\documentclass[11pt, a4paper]{article}
\usepackage{arxiv}
\usepackage[utf8]{inputenc}
\usepackage[sort&compress, round]{natbib}
\usepackage{xcolor}
\definecolor{blendedblue}{rgb}{0.2, 0.2, 0.6}
\usepackage[bookmarks   = true,
            citecolor   = blendedblue,
            colorlinks  = true,
            linkcolor   = blendedblue,
            urlcolor    = blendedblue,
            citecolor   = blendedblue,
            linktocpage = false,
            hyperindex  = true]{hyperref}
\usepackage{url}
\usepackage{doi}
\usepackage{orcidlink}
\usepackage{fontawesome}
\usepackage{lmodern}
\usepackage{booktabs}
\usepackage{amsmath, amssymb, amsthm}

\usepackage[labelsep=period]{caption}

\renewcommand{\headeright}{}

\usepackage{parskip}
\title{Structural Health Monitoring with Functional Data: Two Case Studies}
\renewcommand{\shorttitle}{Structural Health Monitoring with Functional Data: Two Case Studies}
\date{\today}

\author{
	Philipp Wittenberg\orcidlink{0000-0001-7151-8243} \\
        Department of Mathematics and Statistics\\
        School of Economics and Social Sciences\\
        Helmut Schmidt University\\
        Hamburg, Germany\\
	\texttt{pwitten@hsu-hh.de} \\
	\And
	Sven Knoth\orcidlink{0000-0002-9666-5554} \\
        Department of Mathematics and Statistics\\
        School of Economics and Social Sciences\\
        Helmut Schmidt University\\
        Hamburg, Germany\\
	\texttt{knoth@hsu-hh.de}\\
	\And 
	Jan Gertheiss\orcidlink{0000-0001-6777-4746} \\
        Department of Mathematics and Statistics\\
        School of Economics and Social Sciences\\
        Helmut Schmidt University\\
        Hamburg, Germany\\
	\texttt{gertheij@hsu-hh.de} \\
}

\begin{document}	
\maketitle

\begin{abstract}
Structural Health Monitoring (SHM) is increasingly used in civil engineering. One of its main purposes is to detect and assess changes in infrastructure conditions to reduce possible maintenance downtime and increase safety. Ideally, this process should be automated and implemented in real-time. Recent advances in sensor technology facilitate data collection and process automation, resulting in massive data streams. Functional data analysis (FDA) can be used to model and aggregate the data obtained transparently and interpretably. In two real-world case studies of bridges in Germany and Belgium, this paper demonstrates how a function-on-function regression approach, combined with profile monitoring, can be applied to SHM data to adjust sensor/system outputs for environmental-induced variation and detect changes in construction. Specifically, we consider the R package \texttt{funcharts} and discuss some challenges when using this software on real-world SHM data. For instance, we show that pre-smoothing of the data can improve and extend its usability.
\end{abstract}


\section{Introduction}\label{sec:introduction}
Structural Health Monitoring (SHM) typically implements a damage detection strategy for structural and mechanical systems, continuously assessing and monitoring their condition \citep{Farrar.Worden_2012}. Especially with the availability of modern sensor technology, it is an emerging field getting more and more attention. However, a well-known challenge in using SHM methods is dealing with external environmental factors, e.g., temperature, relative humidity, wind speed, and others, and their impact on the state and behavior of the monitored structure. Developing new ways to adjust for these confounding effects is essential. A recent review of the influence of temperature is given in \cite{Han.etal_2021}. 

Statistical process monitoring methods have long been incorporated into SHM practice to detect changes, refer to \cite{Sohn:Czar:Farr:2000}. On the other hand, methods such as profile monitoring, functional data analysis (FDA), and functional principal component analysis (FPCA) are rarely used in SHM  \citep{Momeni.Ebrahimkhanlou_2022}, although they offer a tool-set for holistic and transparent modeling of the vast and complex data structures, which are quite common within the SHM field.

This paper will demonstrate in two case studies how the recently proposed function-on-function regression approach by \cite{Centofanti.etal_2021} can be applied to structural health monitoring data to adjust the system outputs for environmental factors. For this purpose, we utilize the R \citep{R_2023} and \texttt{funcharts} software \citep{Capezza.etal_2023, Capezza.etal_2023c}, and show that pre-smoothing of the data can improve and extend its usability.

The remainder of the paper is structured as follows. Section~\ref{sec:data} introduces the two case studies and data sets. Section~\ref{sec:methods} provides the necessary methodological background of multivariate functional data analysis, function-on-function regression, and associated control charts. In Section~\ref{sec:case_studies}, the \texttt{funcharts} R package is applied to our SHM data, and the results are discussed. Section~\ref{sec:conclusion} concludes with a summary and further discusses the challenges and software limitations noticed with our SHM data.

\section{Data description}\label{sec:data}
\subsection{Steel bowstring railway bridge KW51}
The first data set considers the steel bowstring railway Bridge KW51 in Belgium near Leuven. 
As described in Maes et al., 2022, this bridge had a construction error and had to be subsequently strengthened by reinforcing the connections of the diagonals to the arches and the bridge deck. Researchers from the University of Leuven \citep{Maes.Lombaert_2021} placed, over 15.5 months, several accelerometers measuring vibration characteristics on the arches of the bridge and the bridge deck to collect data before, during, and after the retrofitting. Therefore the data set can be split into the following three parts. The first part, before retrofitting: 10/02/2018--05/15/2019; the second part, during retrofitting: 05/15/2019--09/27/2019; and the third part after retrofitting: 09/27/2019--01/15/2020. Additionally, the bridge's steel temperature, the bridge deck's relative humidity, and several other environmental variables from a nearby weather station located on the Leuven University campus were measured. The bridge is excited by ambient vibrations from trains, wind, or vehicles passing under the bridge, and the accelerometers registered the resulting vibrations. This dynamic behavior of the structure can be represented by a combination of modes characterized by natural frequency, damping ratio, and mode shape, which depend on the structure's specific geometry and material properties. In order to extract the modes and natural frequencies, an operational modal analysis (OMA) was used in one-hour intervals after appropriate pre-processing with filters and other signal processing techniques; for details on OMA, see, e.g., \cite{Rainieri.Fabbrocino_2014}. Fourteen modes and their natural frequencies were identified and subsequently tracked over time. All data in 1-hour intervals is available at \cite{Maes.Lombaert_2020} and more details on the particular procedures are given in \cite{Maes.Lombaert_2021} and \cite{Maes.etal_2022}.
Note that data for some environmental variables is missing over more extended periods. Furthermore, not all 14 modes were always identified in the 1-hour intervals in the OMA throughout the whole period and thus could not be tracked over time. Therefore, there are many missing values present in the available data set.

\begin{figure}[!htb]
\includegraphics[width=.505\linewidth]{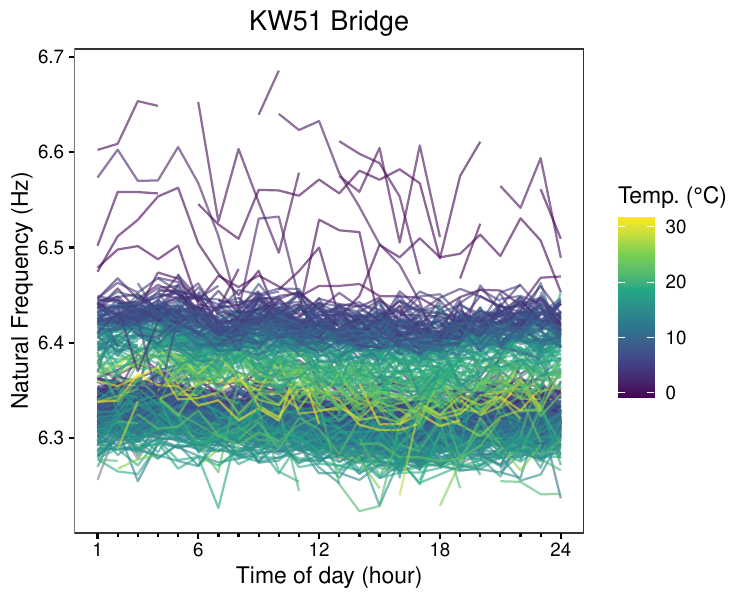}
\includegraphics[width=.505\linewidth]{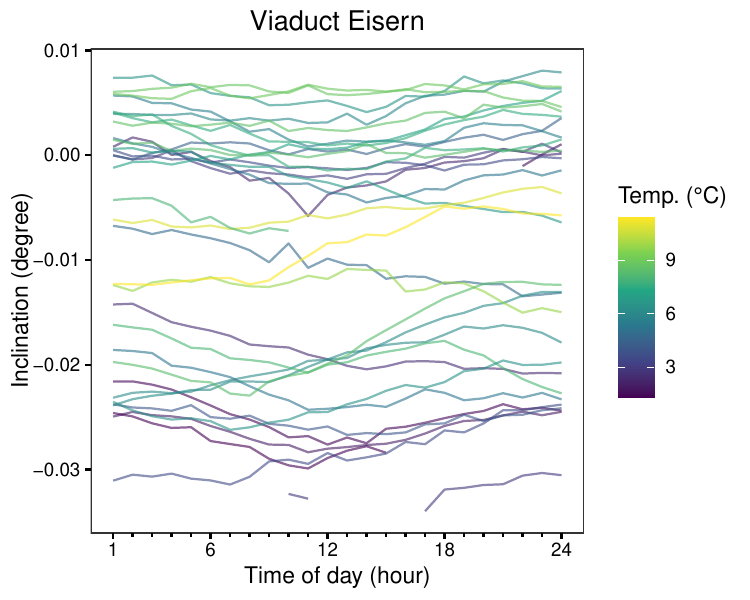}
\caption{Functional profiles of two bridge data sets. The left panel, KW51 bridge, shows the natural frequency of Mode 13 and the average daily temperature. The right panel, viaduct Eisern, shows inclination of a sensor on the first floor of the pillar and the average daily temperature.}
\label{fig:raw_bridge_data}
\end{figure}

We will consider the dynamic response variable natural frequency of Mode 13 and two environmental variables (steel temperature and relative humidity) when studying the bridge's change in condition during retrofitting in Section~\ref{sec:KW51_cases_study}. The left panel of Figure \ref{fig:raw_bridge_data} shows the natural frequency of mode 13, interpreted as functional profiles per day, and their inverse relationship to the average daily temperature. 
The figure also shows the partially missing data and some outlying data points of the natural frequency (observed between 01/18/2019 and 02/02/2019).

\subsection{Highway viaduct Eisern}
The second data set, new and unpublished so far, is from the prestressed concrete viaduct ``Eisern'' on Highway A45, near Siegen, located in the federal state of North Rhine Westfalia in Germany. The bridge was scheduled for demolition and a rebuild to extend its traffic capacity. Researchers from the civil engineering and statistics department of Helmut Schmidt University were given a permit to install sensors during preparation for the demolition scheduled in spring 2023. This opportunity allowed for capturing on-site sensor data in three phases of a fieldwork study. Phase 1 was with regular traffic on the bridge. In phase 2, the bridge was closed to traffic, with moderate construction work around the bridge site and some deconstruction at the superstructure. In phase 3, particular bridge elements (supporting pillars) were pre-weakened before the actual demolition at the end of March. An official construction site notebook kept detailed information about scheduled and performed construction processes.

The data was collected between 02/08/2023 and 03/20/2023, with interruptions due to a blackout at the construction site and reconfiguring sensor placement between 03/01/2023 and 03/8/2023. Therefore, for some days, only partial data were collected. 

Inclinometers and accelerometers were placed at two locations on the bridge and measurements were taken using two independent systems, one inside the superstructure and the other at different levels inside one of the supporting pillars. In what follows, we will only consider the output of one uniaxial inclinometer (INC) on the first floor inside the support pillar structure that measures the inclination in the transverse direction (in degrees) using the sampling rate of 100~Hz. In addition, environmental data were recorded by a weather station placed on the ground floor inside the pillar. At 1~Hz sampling rate, the temperature (temp) measured in °C and relative humidity (humid) measured in \% are recorded. A detailed data description paper for the viaduct Eisern is currently under preparation. Based on the recommendation of \citet{Han.etal_2021}, static response data used to analyze damage detection under the influence of slowly changing temperatures can be further aggregated. Accordingly, with their high sampling rates, the inclination and environmental data were averaged to one-hour time intervals utilizing the median. These data are then used in the case study in Section~\ref{sec:Eisern_case_study}. The right panel of Figure~\ref{fig:raw_bridge_data} shows the 41 processed daily inclination profiles with mostly 24 observations per profile and partially missing data. For illustration, the hourly temperature values were again averaged to daily values to emphasize the association between inclination and temperature. 

In summary, both data sets, the steel arch bridge KW51 and the prestressed concrete bridge Eisern, can be interpreted as functional data. Although they consider dynamic and static responses and differ in the temperature range, a clear association with the temperature is identified. Therefore, the following section revisits functional principal component analysis, function-on-function regression, and multivariate control charts for confounder-adjusted system outputs and further evaluation of changes in structural conditions.

\section{Methods}\label{sec:methods}
We briefly describe multivariate functional principal component analysis (MFPCA) and the functional regression control chart (FRCC) framework under the presence of functional covariates proposed in \cite{Centofanti.etal_2021}, which will be used on our SHM data in Section~\ref{sec:case_studies}. Throughout the paper, we use a very broad definition of `confounder' or `confounding variable' as ``a variable whose presence affects the variables being studied so that the results do not reflect the actual relationship''~\citep{Pourhoseingholi.etal_2012}. Specifically, the variables being studied are the status of the bridge and the sensor/system outputs, whereas the environmental factors are the (potential) confounders. In contrast to the standard definition, confounders, in our case, mainly affect the behavior of the bridge being studied and, hence, the sensor/system outputs, but not necessarily the bridge's condition. Nevertheless, they can obscure the relationship between those two, making it much harder to use sensor data to monitor the bridge's condition. For instance, low temperatures may affect system outputs in a way that would indicate damage if temperatures were mild. In other words, ignoring environmental influences may lead to wrong conclusions (false alarms or false negatives). To account for these confounding effects, we use the regression approach. That means using the confounders as covariates in a regression model to adjust for their effects on the system outputs. The FRCC approach then consists of three steps: (1) defining the functional regression model (response and covariates), (2) the method for model fitting, and (3) the monitoring strategy of the functional residuals.

\subsection{Multivariate functional principal component analysis}\label{sec:methods:MFPCA}
MFPCA is the fundamental building block of the monitoring framework implemented in \texttt{funcharts}. MFPCA is based on the multivariate version of the Karhunen-Loéve Expansion. In short, we assume that we have multivariate functional random variables $\mathbf{X}(t) = (X_1(t), \ldots, X_p(t))^\top \in \mathbb{R}^p$ in a Hilbert space, with $t \in \mathcal{T}$, $\mathcal{T}$ a compact interval in $\mathbb{R}$, and all components $X_k(t)$, $k=1,\ldots,p$, are square integrable, one-dimensional (random) functions with mean $\text{E}(\mathbf{X}(t)) = \mathbf{0}$, for all $t \in \mathcal{T}$. In general, $X_k$ may be observed on different (dimensional) domains, but in our case, the domain is always the same, namely one day, i.e., an interval of 24 hours. Then, under mild conditions, it holds that \citep{Happ.Greven_2018}

\begin{equation}
    \text{Cov}(X_k(t),X_k(s)) = \sum_{l=1}^\infty \nu_l \psi_{lk}(t)\psi_{lk}(s), \;\; t,s \in \mathcal{T},
\end{equation}

with orthonormal $\boldsymbol{\psi}_l(t) = (\psi_{l1}(t),\ldots,\psi_{lp}(t))^\top$, and

\begin{equation}\label{eq:MKLE}
    \mathbf{X}(t) = \sum_{l=1}^\infty \xi_l \boldsymbol{\psi}_l(t), \;\; t \in \mathcal{T},
\end{equation}

where $\xi_l$ are mean zero, pairwise uncorrelated random variables with variance $\nu_l$, also compare \citet{KonSta:2023}. A finite, $L$-dimensional approximation of $\mathbf{X}(t)$ is then obtained through

\begin{equation}\label{eq:approX}
    \hat{\mathbf{X}}^L(t) = \sum_{l=1}^L \xi_l \boldsymbol{\psi}_l(t).
\end{equation}

The functions $\boldsymbol{\psi}_l(t)$ are called (multivariate) functional principal components and can be estimated through eigendecompositions of appropriately chosen empirical covariance matrices calculated from a sample $\mathbf{X}_1,\ldots,\mathbf{X}_n$, where for each $\mathbf{X}_i$ equation \eqref{eq:MKLE} holds, i.e., $\mathbf{X}_i(t) = \sum_{l=1}^\infty \xi_{il} \boldsymbol{\psi}_l(t)$, and the FPC scores $\xi_{il}$ are independent over $i$; see \citet{Ramsay.Silvermann_2005} and \citet{Happ.Greven_2018} for details and ways to estimate the so-called scores $\xi_l$. Specifically, \citet{Capezza.etal_2023c} use $\hat\xi_l = \sum_{k=1}^p \int_\mathcal{T} X_k(t)\hat\psi_{lk}(t) dt$.

In what follows, we assume that we have corresponding $M$- and $L$-dimensional approximations of the functional data to be monitored $Y(t)$ and the (functional) covariates $X_{1}(t),\ldots,X_p(t)$, respectively. Given all functional data have been standardized (empirically) to have zero mean (which will be assumed throughout the paper), this means, for day $i$, we have

\begin{equation}\label{eq:approYM} \hat{Y}_i^M(t)=\sum_{m=1}^M\xi_{im}^Y\Psi_{m}^Y(t), \quad t\in \mathcal{T},
\end{equation}
and
\begin{equation}\label{eq:approXL}
\hat{X}_{ik}^L(s)=\sum_{l=1}^L\xi_{il}^X\Psi_{lk}^X(t), \quad k=1,\dots,p, \quad t\in \mathcal{T},
\end{equation}
respectively. Note that in practice, both the eigenfunctions and the scores are not directly observable, which means that the corresponding quantities in \eqref{eq:approYM} and \eqref{eq:approXL} have to be replaced by estimates $\hat\Psi_{m}^Y$, $\hat\Psi_{l}^X$, and $\hat\xi_{im}^Y$, $\hat\xi_{il}^X$, respectively. The approximations \eqref{eq:approYM} and \eqref{eq:approXL} are the basis for both the function-on-function regression implemented in \texttt{funcharts} to account for potential confounders as well as the control charts used for monitoring. The number of principal components ($M$, resp.~$L$) is typically chosen such that a certain percentage of the total variation in the data, such as 95\% or 99\%, is explained~\citep{Capezza.etal_2023c}. A potential issue when using \texttt{funcharts} to perform MFPCA is that \texttt{funcharts} employs the corresponding \texttt{fda}~\citep{Ramsey.etal_2022} implementation, which is not able to handle missing values in the functional data as often observed with SHM sensor data (see Section~\ref{sec:data}). For ways to handle this problem, see Section~\ref{sec:case_studies} and \ref{sec:conclusion}. Finally, note that when using \texttt{funcharts}, the functional data are not only centered but standardized by dividing (pointwise) through the respective (empirical) standard deviation function.

\subsection{Function-on-function regression}\label{sec:methods:regr}
To adjust for the covariates, the general function-on-function regression approach used in \texttt{funcharts} is
\begin{equation}\label{eq:general_model}
    Y_i(t)=\beta_0(t) + \sum_{k=1}^p \int_{\mathcal{S}} X_{ik}(s)\beta_k(s,t)ds + \epsilon_i(t), \quad i=1,\dots,n, \quad t\in \mathcal{T}.
\end{equation}
If all functional data have been standardized as assumed above, the functional intercept $\beta_0(t)$ can be omitted. Model~\eqref{eq:general_model} is very popular in functional data analysis, although it is not the only one available for function-on-function(s) regression; see, e.g., \citet{Scheipl.etal_2016} for alternatives. An important advantage of using the linear model~\eqref{eq:general_model}, however, is that after approximating both the response and the covariates through MFPCA as done above, the coefficient functions $\beta_k(s,t)$ take the form
\begin{equation}
    \beta_k(s,t) = \sum_{l=1}^L\sum_{m=1}^Mb_{lm}\Psi_{lk}^X(s)\Psi_m^Y(t), \quad s \in \mathcal{S}, \quad t \in \mathcal{T}, \quad k=1,\dots,p,
\end{equation}
and the functional model~\eqref{eq:general_model} turns into a multivariate, yet scalar model
\begin{equation}
    \xi_{im}^Y=\sum_{l=1}^L\xi_{il}^X b_{lm} + \epsilon_{im}, \quad m=1,\dots,M.
\end{equation}
Since the predictor scores are uncorrelated, basis coefficients $b_{lm}$ can be estimated through
\begin{equation}
    \hat{b}_{lm} = \frac{\sum_{i=1}^n(\hat\xi_{im}^Y\hat\xi_{il}^X)}{\sum_{i=1}^n(\hat\xi_{il}^X)^2}
\end{equation}
and plugged-in to obtain the estimate $\hat{\beta}_k(s,t)$. The functional response given the covariates can then be predicted by
\begin{equation}\label{eq:predY}
\hat{Y}_i(t)=\sum_{k=1}^p\int_\mathcal{S}X_{ik}(s)
\hat{\beta}_k(s,t)ds, \quad t \in \mathcal{T}
\end{equation}
and functional residuals are obtained as
\begin{equation}
    e_i(t)=Y_i(t)-\hat{Y}_i(t), \quad t \in \mathcal{T}.
\end{equation}
Given all unknown quantities (parameters, scores, functions) have been estimated on in-control samples $i=1,\ldots,n$ only, \eqref{eq:predY} is an estimate of what profile would be expected, conditioned on the covariates, if the process is still in control. Consequently, the residuals of future observations (which have not been used for model fitting) can be used for monitoring.

\subsection{Multivariate control charts}\label{sec:methods:ccharts}
For monitoring $e_i(t)$, this process is first approximated through (univariate) $R$-dimensional FPCA, again with all crucial quantities being estimated on in-control samples $i=1,\ldots,n$ only, resulting in (estimated) eigenfunctions $\hat\psi_r^{e}(t)$, eigenvalues/variances $\hat\nu_r^{e}$, and scores $\hat\xi_{ir}^{e}$. Then the day-specific
Hotelling's statistic $T_i^2$ is defined as
\begin{equation}\label{eq:controlchart_T2}
    T^2_i=\sum_{r=1}^R (\hat\xi_{ir}^{e})^2/\hat\nu_r^{e}
\end{equation}
and the Squared Prediction Error (SPE) is defined as
\begin{equation}\label{eq:controlchart_SPE}
    \text{SPE}_i= \int_\mathcal{T}(e_{i}(t)-\hat{e}_{i}(t))^2dt,
\end{equation}
where $\hat{e}_i(t) = \sum_{r=1}^R \hat\xi_{ir}^{e}\hat\psi_r^{e}(t)$ is the $R$-dimensional FPCA-based approximation of $e_i(t)$. For monitoring future observations, only the scores need to be calculated (using the eigenfunctions estimated on the in-control samples), and control charts can be constructed analogously to \eqref{eq:controlchart_T2} and \eqref{eq:controlchart_SPE}. The control limits are obtained by estimating the empirical quantiles using kernel density estimates with Gaussian kernel on the in-control data from \eqref{eq:controlchart_T2} and \eqref{eq:controlchart_SPE}  \citep{Centofanti.etal_2021}. For future observations, an alarm is signaled if one of the statistics exceeds its own control limit.

\section{Case studies}\label{sec:case_studies}
\subsection{KW51 bridge}\label{sec:KW51_cases_study}
\begin{figure}[!htb]
\centering
\includegraphics[width=.8\linewidth]{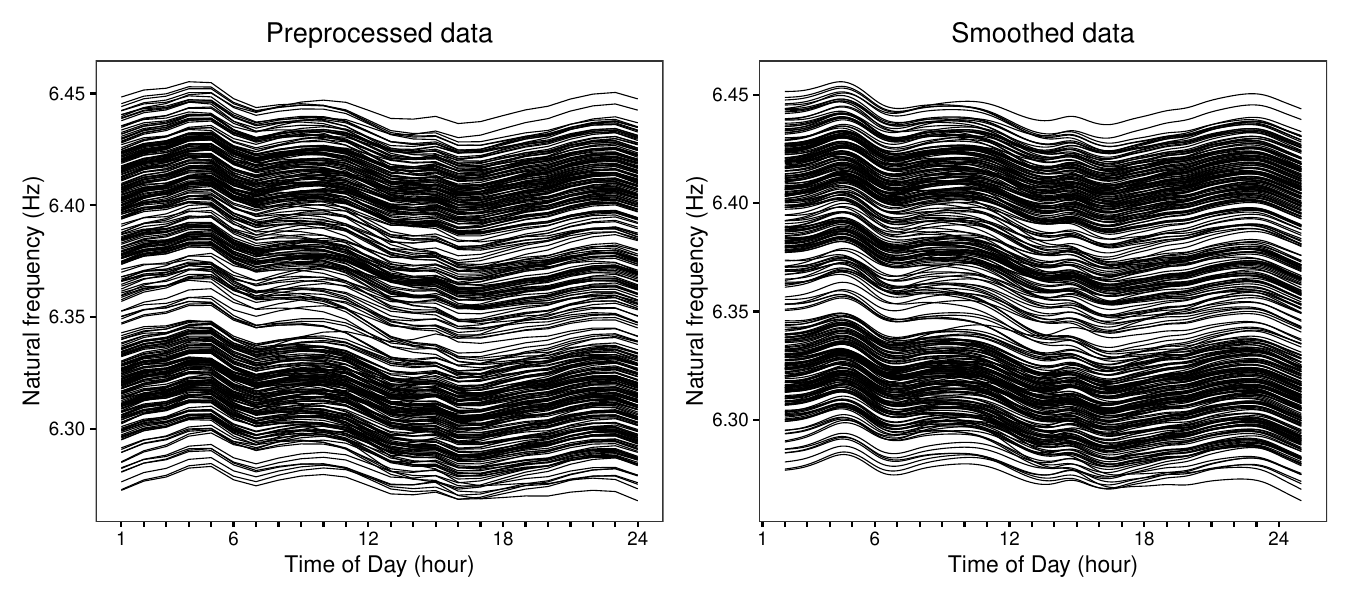}
\caption{Preprocessed (\texttt{refund}) and subsequently smoothed (\texttt{funcharts}) data for the KW51 bridge response variable.}
\label{fig:response_raw_smooth_KW51}
\end{figure}

\begin{figure}[!htb]
\centering
\includegraphics[width=.8\linewidth]{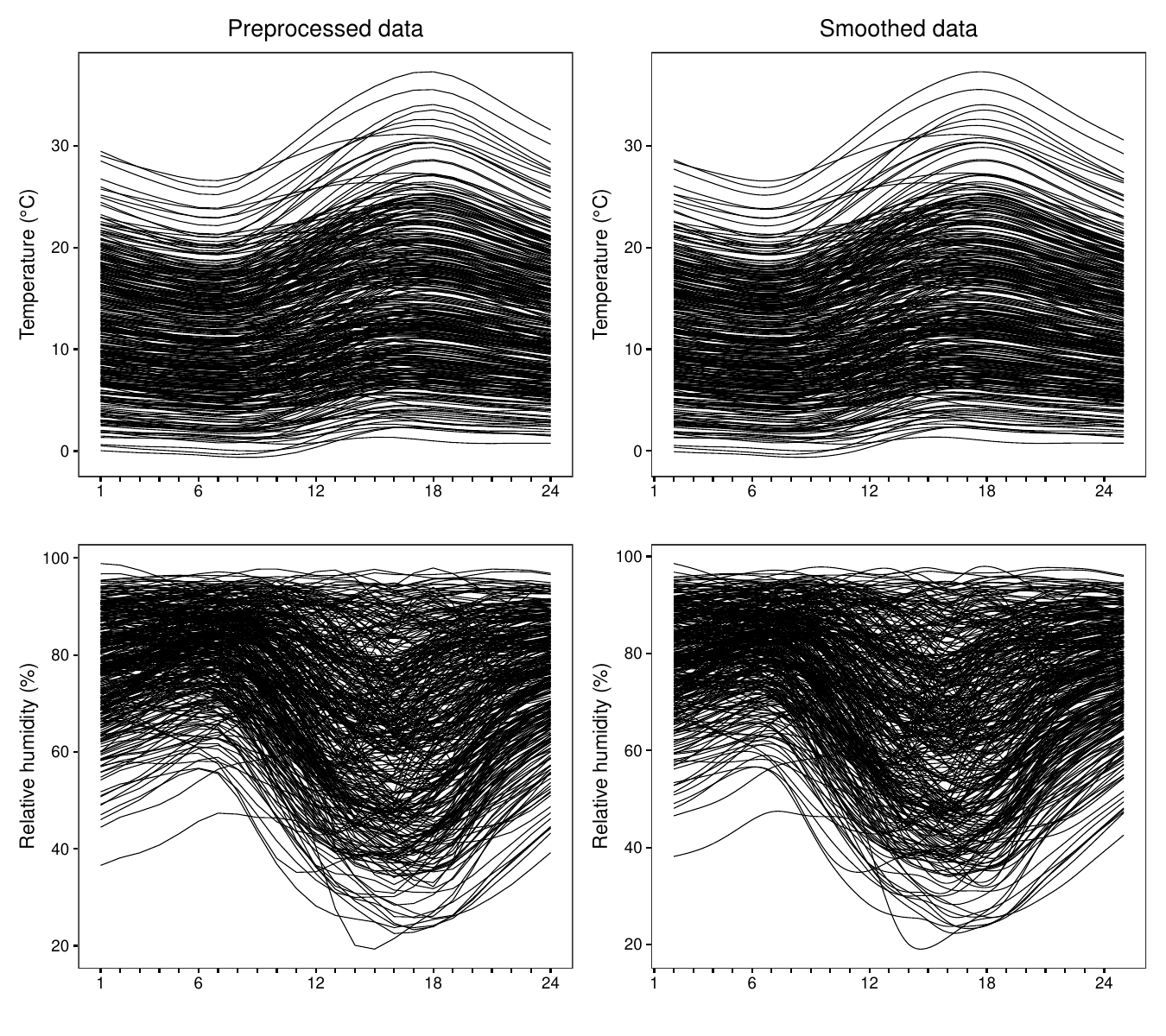}
\caption{Preprocessed (\texttt{refund}) and subsequently smoothed (\texttt{funcharts}) data for the KW51 bridge covariates temperature (upper
row) and relative humidity (lower row).}
\label{fig:covariate_raw_smooth_KW51}
\end{figure}

A major obstacle we faced in this particular case study was the number of missing values in the initial data. Missing data is not unusual in SHM. It can be related to sensor change, sensor downtime, a blackout of the measurement system,  modes that could not be identified during operational modal analysis and their tracking over time. The total number of days in the study was 472, of which only 102 profiles (21.6\%) were completely observed without missing values and directly usable by \texttt{funcharts}. Of the 472 profiles, 94 were excluded from the study due to fewer than four observations per profile. These missings were mainly caused by missing data periods for the covariates and the fact that outliers between day 109 and 124, as mentioned in Section \ref{sec:data}, were removed. After preprocessing, the number of complete profiles increased to 364, which means that 262 profiles (55.5\%) were recovered. The missing data were recovered through pre-smoothing using the \texttt{fpca.face()} function of the \texttt{refund} R-package \citep{Xiao.etal_2013, Goldsmith.etal_2022}. Functional principal component analysis (FPCA) is also a popular tool for predicting/reconstructing the underlying profiles of sparsely observed functional data; compare, e.g., \citet{Yao.etal_2005, Di.etal_2014}. Profiles of the preprocessed data using \texttt{refund} and subsequently imported and smoothed data using \texttt{funcharts}' \texttt{get\_mfd\_df()} function (with default values where the number of B-spline basis functions of order four is set to 30 with equally spaced knots and the smoothing penalty is chosen based on penalizing the integrated squared second derivative) for the response and covariate data are shown in Figure \ref{fig:response_raw_smooth_KW51} and Figure \ref{fig:covariate_raw_smooth_KW51}, respectively. Both the preprocessed and subsequently smoothed data look very similar. This is not surprising. First-stage FPCA already smoothed the data, so additional smoothing essentially reproduces the (already smooth) data. Of course, first applying univariate FPCA to recover the curves/impute the missing values is just a ``quick fix'' for making \texttt{funcharts} applicable. It would be desirable to carry out data recovery, smoothing, and MFPCA (as described above) simultaneously; see also Section \ref{sec:conclusion}.

Before setting up the control charts, the MFPCA implemented in the \texttt{fof\_pc()} function in \texttt{funcharts} was performed (on the reconstructed/smoothed data) using the methods described in Section~\ref{sec:methods:MFPCA}. Subsequently, MFPCA results from both the multivariate functional covariates and the response were used to regress the response scores on the scores of the covariates using a linear regression model, as described in Section~\ref{sec:methods:regr}. Finally, FPCA of the error process led to the control charts (compare Section~\ref{sec:methods:ccharts}). A value of 95\% explained total variance is chosen for each of the three (M)FPCAs performed, leading to $M = 2, L = 4$, and $R = 2$.

\begin{figure}[!htb]
\centering
\includegraphics[width=.7\linewidth]{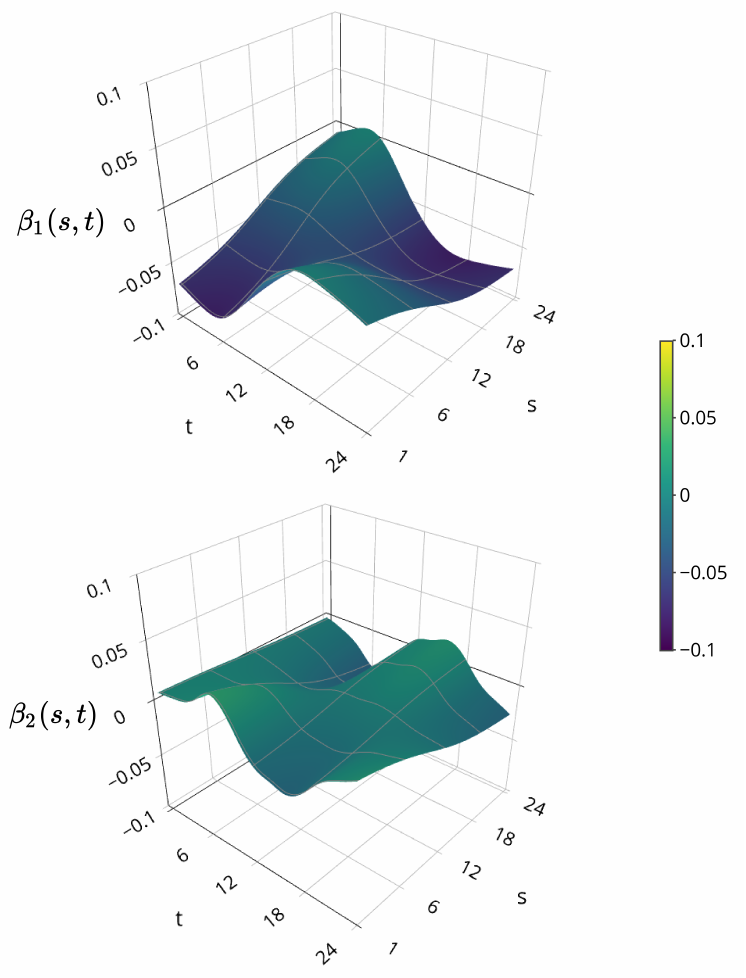}
\caption{Covariate effect $\beta_1(s,t)$ of temperature (top) and $\beta_2(s,t)$ of humidity (bottom) in the function-on-function regression model~\eqref{eq:general_model} for the KW51 bridge.}
\label{fig:effects_KW51}
\end{figure}

In addition to temperature, we use relative humidity as a second covariate in our analyses, as this can also influence the response behavior of the bridge \citep{Keshmiry.etal_2023}. Figure~\ref{fig:effects_KW51} shows the estimated coefficient surfaces $\hat\beta_1(s,t)$ and $\hat\beta_2(s,t)$ for temperature (left) and rel. humidity (right), respectively (using the function \texttt{plot\_bifd()}). Apparently, the temperature has a negative effect on the 13th natural frequency of KW51 (mode 13) as considered here. Such negative effects of temperature on natural frequencies have also been found in other studies, see \cite{Han.etal_2021}. As seen in Figure~\ref{fig:effects_KW51}~(left), this effect is most pronounced along the diagonal, meaning that concurrent temperature is most relevant. The effect of humidity (Figure~\ref{fig:effects_KW51}, right), by contrast, is rather weak and hard to interpret.

As the KW51 bridge has exact dates for the time before, during, and after the retrofitting, we set the end of phase I to 15 days before the retrofitting started (30th of April 2019), corresponding to profile 106 in the compiled data set. The control charts are set up using function \texttt{regr\_cc\_fof()}. Complete phase I data was used for estimating the parameters and control limits. The overall type-I error probability of 1/370.4 using a Bonferroni correction was applied to both charts.
\begin{figure}[!htb]
\centering
\includegraphics[width=.9\linewidth]{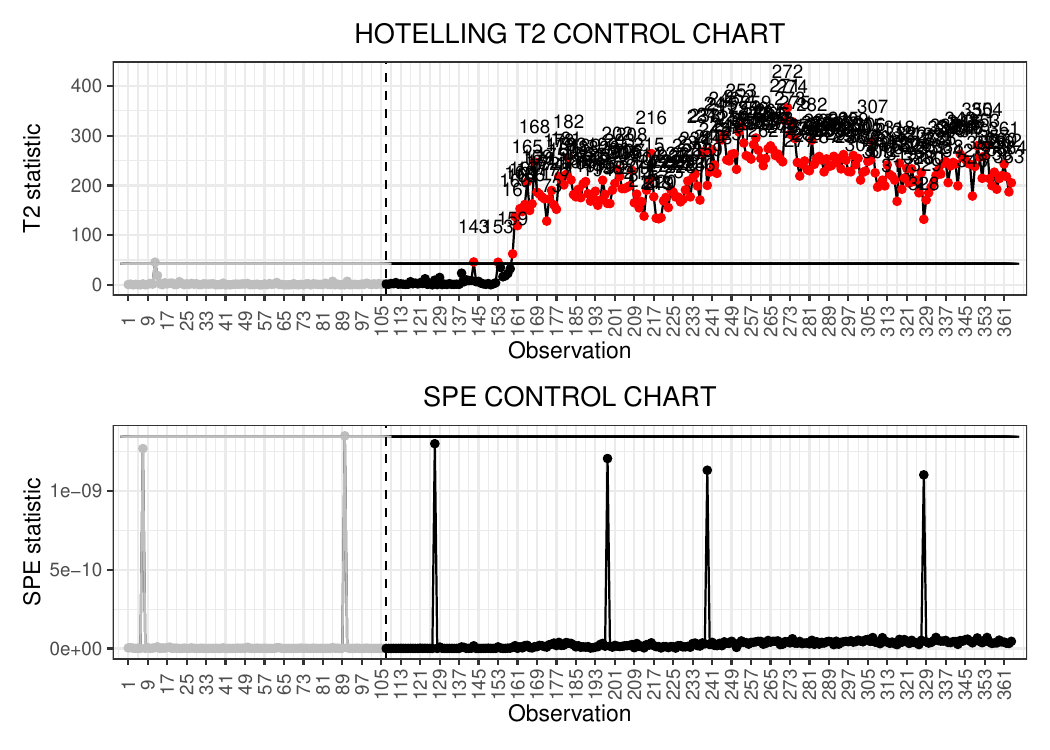}
\caption{$T^2$ and SPE control charts for inclination sensor at the KW51 railroad bridge.}
\label{fig:control_charts_KW51}
\end{figure}
The two control charts according to equation \eqref{eq:controlchart_T2} and \eqref{eq:controlchart_SPE} are shown in Figure \ref{fig:control_charts_KW51}. Each signalled one alarm in the in-control phase. The Hotelling at observation 12 and the SPE chart at observation 90. The Hotelling chart detects the first out-of-control signal at observation 143, followed by a sequence of signals ten observations later. The actual retrofitting started at observation 121, corresponding to the 15th of May 2019. It can be noted that the Hotelling chart issues signals and detects the shift compared to the SPE control chart, which issues no signals. We can conclude that the FPCA approximation, i.e., the found eigenfunctions, still work well for the retrofitting phase, and the corresponding scores clearly indicate the change in the bridge's behaviour through the Hotelling chart \eqref{eq:controlchart_T2}.  

\subsection{Viaduct Eisern}\label{sec:Eisern_case_study}
\begin{figure}[!htb]
\centering
\includegraphics[width=.8\linewidth]{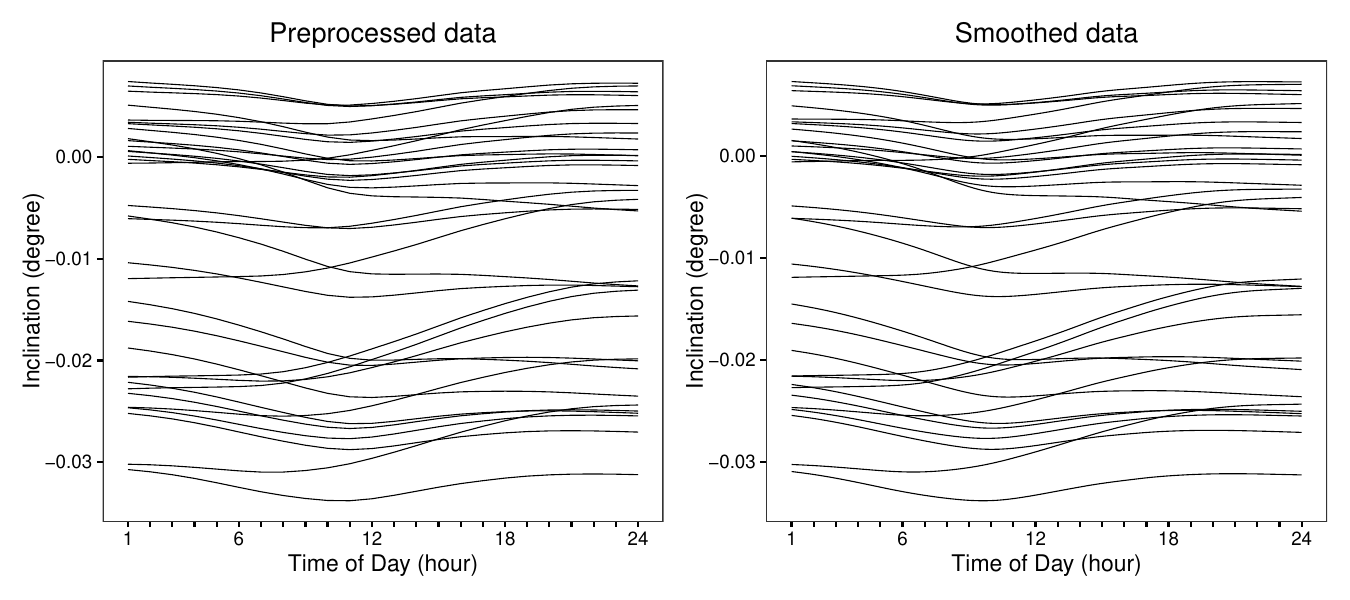}
\caption{Preprocessed (\texttt{refund}) and subsequently smoothed (\texttt{funcharts}) data for the viaduct Eisern response variable.}
\label{fig:response_raw_smooth_EISERN}
\end{figure}

\begin{figure}[!htb]
\centering
\includegraphics[width=.8\linewidth]{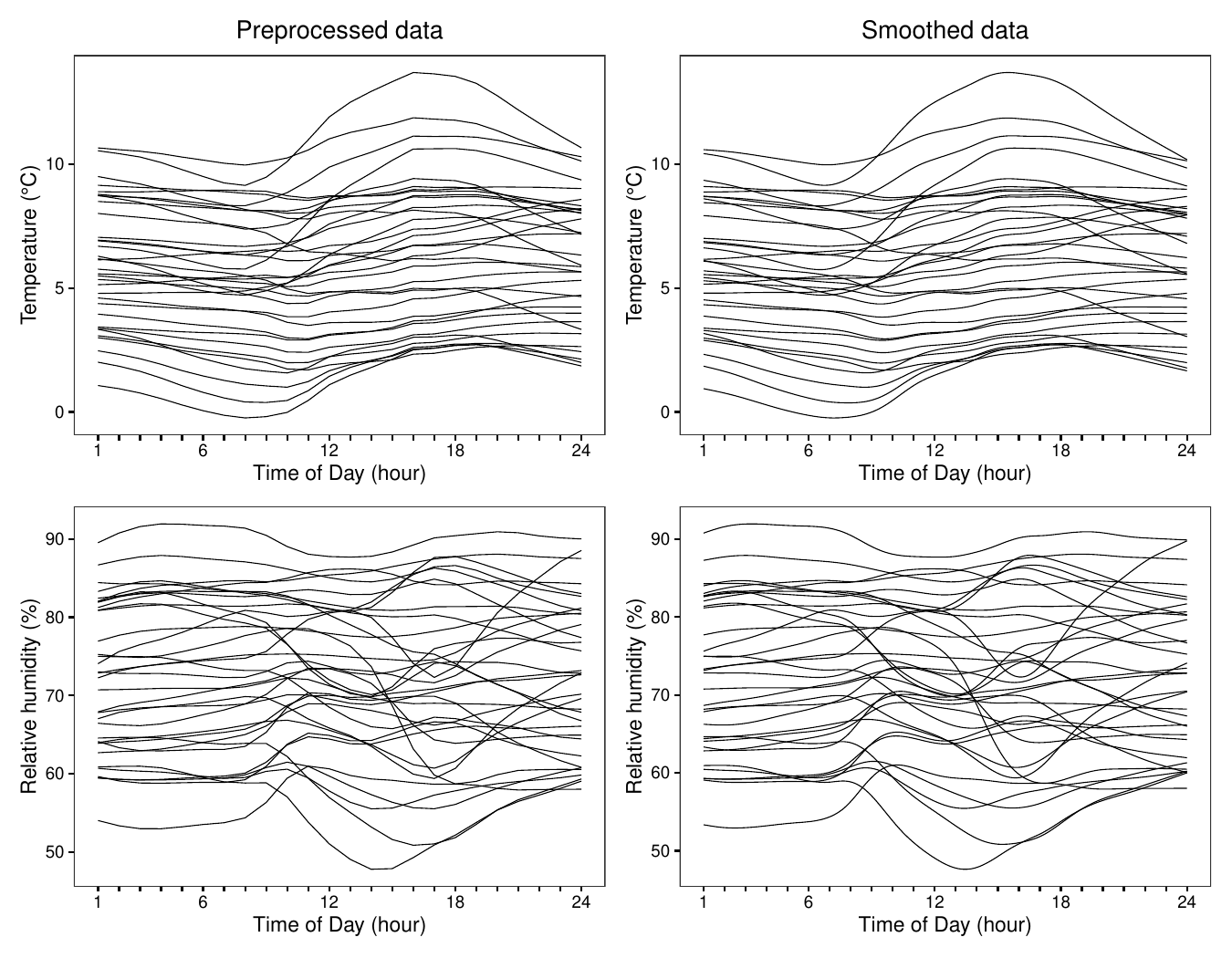}
\caption{Preprocessed (\texttt{refund}) and subsequently smoothed (\texttt{funcharts}) data for the viaduct Eisern covariates temperature (upper row) and relative humidity (lower row).}
\label{fig:covariates_raw_smooth_EISERN}
\end{figure}

The initial study data included 41 days. Seven profiles (days 1, 23-28), had less than four observations each. Four profiles (22, 29, and 41), recorded at least four observations but were still incomplete, with their proportion of missing values ranging from 0.375.5 to 0.583. Thus, the proportion of profiles with missing values in the data set is 20.5\%. As mentioned in \cite{Jensen.etal_2008}, it would be possible, but not recommended, to fit curves to profiles with a very limited number of observations and then use them in monitoring. Therefore, we decided to exclude profiles with fewer than four observations per profile from further analysis. This reduces the data to 34 profiles, including four profiles with partially missing data. As already mentioned, fully observed data must be available to apply the \texttt{funcharts} package. Hence, as done before, the missing data in the inclination, temperature, and humidity profiles were supplemented with a pre-smoothing procedure using the \texttt{fpca.face()} function of the \texttt{refund} R-package. The preprocessed and completed data can now be utilized by \texttt{funcharts}. As before, we used the function \texttt{get\_mfd\_df()} to generate functional data objects with default values.
Figure \ref{fig:response_raw_smooth_EISERN} and \ref{fig:covariates_raw_smooth_EISERN} show the preprocessed and smoothed data for the response and the covariates, respectively. Similarly to the first case study, one can see that both the preprocessed and smoothed data look very similar.

Next, as in the previous case study, we applied the function \texttt{fof\_pc()} to perform the MFPCA with $L=2, M=2$, and $R=2$. Recall that the bridge data is split into three phases. With regular traffic on the bridge, the first 14 days are used as a short Phase I to estimate the control limits and parameters; the second phase corresponds to moderate construction work on superstructure demolition, and the third one to drilling holes in the pillar and pre-weakening by cutting triangles into the supporting pillar (03/14/2023--03/16/2023, corresponding to observation days 28--30).

\begin{figure}[!htb]
\centering
\includegraphics[width=.7\linewidth]{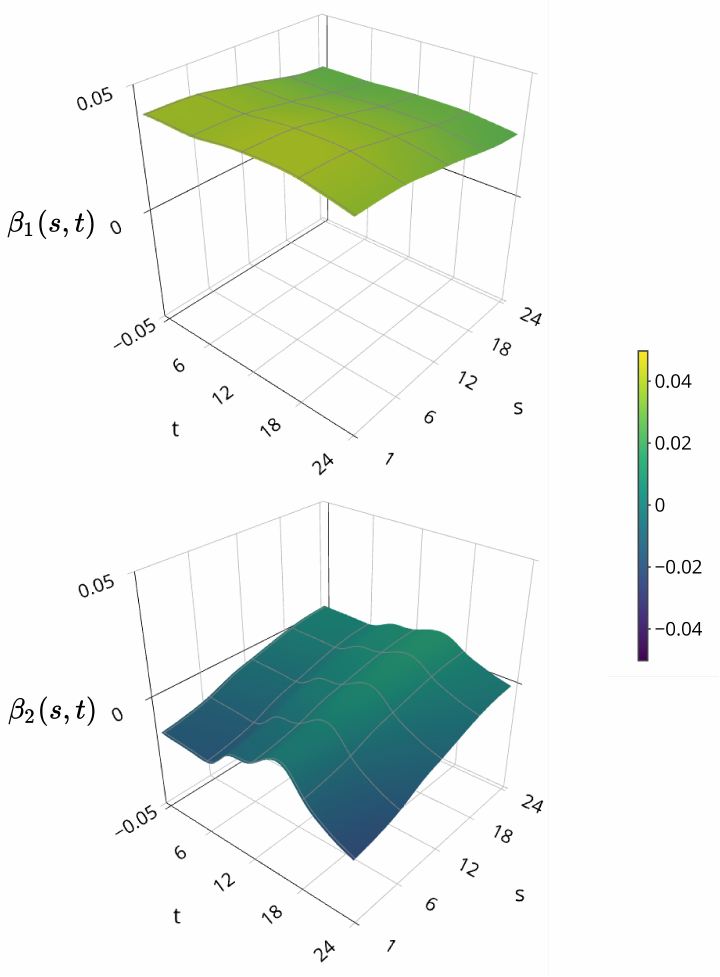}
\caption{Covariate effect $\beta_1(s,t)$ of temperature (top) and $\beta_2(s,t)$ of humidity (bottom) in the function-on-function regression model~\eqref{eq:general_model} for the Eisern viaduct.}
\label{fig:effects_Eisern}
\end{figure}

Figure \ref{fig:effects_Eisern} shows the covariate effects in the function-on-function regression model for the viaduct Eisern based on the Phase I data. It is visible that temperature (top panel) has a positive effect, while the influence of relative humidity (bottom panel) is low to negligible. The positive relationship between temperature and static responses is also found in the literature; see \cite{Han.etal_2021} and the references therein. More specifically, however, the left plot shows a very flat, implausible coefficient function. According to the function shown, future temperature (e.g., around hour 20) affects past inclination (e.g., at hour 10).

\begin{figure}[!htb]
\centering
\includegraphics[width=.9\linewidth]{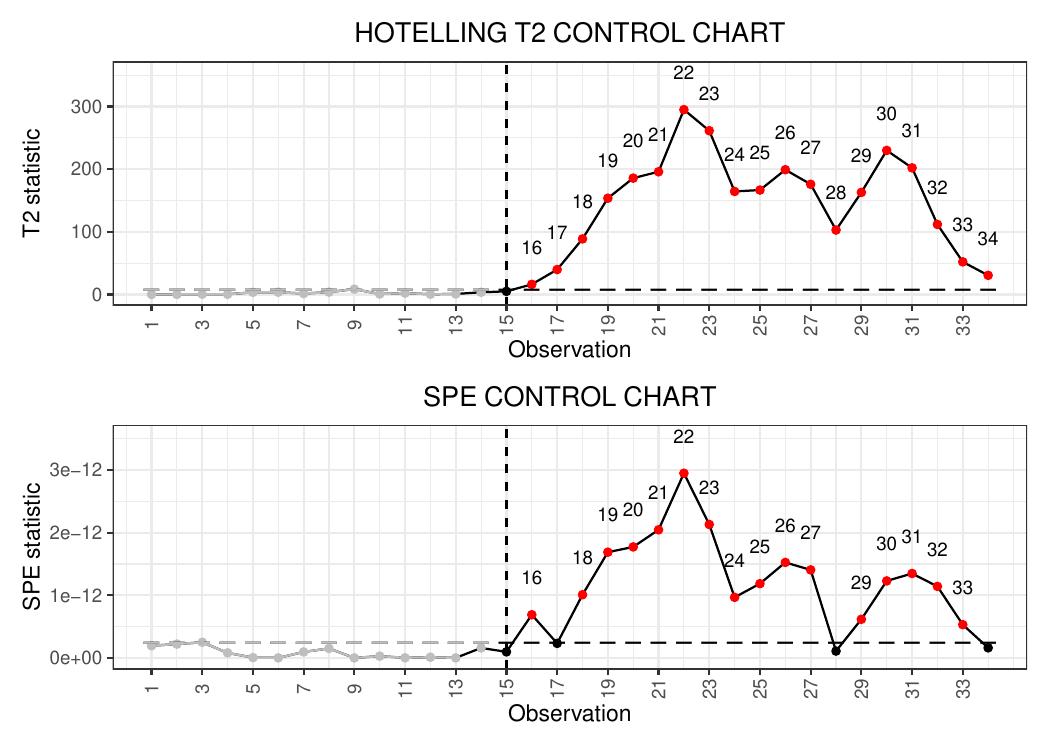}
\caption{$T^2$ and SPE control charts for inclination sensor at the Eisern viaduct.}
\label{fig:control_chart_Eisern}
\end{figure}

Figure \ref{fig:control_chart_Eisern} shows the two control charts. The overall type-I error probability is again set to 1/370.4 with Bonferroni correction to both charts. Two alarms in the in-control phase (9, 3) are presumably false alarms caused by some construction work around the pillar. Interpreting the observed alarm points should be done with care; as mentioned earlier, there is a larger interval of ten days with no data between observations 21 and 22.
While the Hotelling chart shows a signal except for the first day of Phase II, there are four days indicated as ``in-control'' in the SPE-based control chart. However, both charts give strong signals in the monitoring phase.

\section{Summary and Discussion}\label{sec:conclusion}
This paper showed that/how the function-on-function approach proposed by \cite{Centofanti.etal_2021} can be adapted and applied to adjust the system/sensor output in SHM for environmental confounders. The usability of the software package \texttt{funcharts} implementing the FRCC method was tested in two case studies. Reading the data, handling functional objects, and plotting control charts was straightforward. However, since the three-dimensional effect plots of the \texttt{funcharts} package could not be modified sufficiently, functions of the \texttt{plotly} package \citep{Sievert_2020} were used instead. Furthermore, the results found in both case studies align with the SHM literature. 

However, we also identified some challenges regarding the substantial amount of missing data present in our data sets. Due to the employed \texttt{fda}~\citep{Ramsey.etal_2022} methods, the use of \texttt{funcharts} is restricted substantially. In \cite{Capezza.etal_2023c}, it is stated that ``\textit{\dots the funcharts package assumes that these functional data are densely observed on a set of discrete grid
points\dots }''. As shown in Section~\ref{sec:case_studies}, a preprocessing step before the main analysis, which includes the pre-smoothing of the data to impute the missing values within the profiles, is a quick solution to bypass this restriction. A better solution, however, would be implementing and adapting the MFPCA methodology proposed by \cite{Happ.Greven_2018}, available in the R-packages \texttt{funData} \citep{Happ-Kurz_2020} and \texttt{MFPCA} \citep{Happ-Kurz_2022}. The resulting modification of the \texttt{funcharts} would then only replace the FPCA implementation of the \texttt{fda} package by utilizing the \texttt{MFPCA} and \texttt{funData} packages avoiding the additional pre-smoothing step. By doing so, data reconstruction (if necessary), smoothing, and MFPCA would be carried out simultaneously (not successively), making \texttt{funcharts} also applicable to sparse and irregularly sampled functional data.

Another issue concerns the functional regression model~\eqref{eq:general_model} to account for covariate effects. First, by construction, only a linear relationship between the functional response and the covariates can be modeled. However, as indicated in the SHM literature (see \cite{Han.etal_2021} and references therein), some environmental covariates, such as temperature, may affect the natural frequencies of the dynamic response variables in a nonlinear way.
In addition, by integrating over the entire domain $\mathcal{S}$ in \eqref{eq:general_model}, the model allows future covariate values to influence the response's current value. This can lead to problems when interpreting the results, as seen with the viaduct Eisern in Section~\ref{sec:case_studies}. Potential modifications to resolve these problems could be \emph{concurrent}, \emph{nonlinear}, and/or \emph{historical} functional effects in model~\eqref{eq:general_model}; see, e.g., \citet{Scheipl.etal_2016} for details. 

\section*{Acknowledgement}\label{sec:Acknowledgement}
This research is funded by dtec.bw -- Digitalization and Technology Research Center of the Bundeswehr. dtec.bw is funded by the European Union -- NextGenerationEU. The authors thank all Viaduct Eisern data collection team members for granting access to the data.

\bibliographystyle{abbrvnat}
\bibliography{references.bib}

\begin{thebibliography}{27}
\providecommand{\natexlab}[1]{#1}
\providecommand{\url}[1]{\texttt{#1}}
\expandafter\ifx\csname urlstyle\endcsname\relax
  \providecommand{\doi}[1]{doi: #1}\else
  \providecommand{\doi}{doi: \begingroup \urlstyle{rm}\Url}\fi

\bibitem[Capezza et~al.(2023{\natexlab{a}})Capezza, Centofanti, Lepore,
  Menafoglio, Palumbo, and Vantini]{Capezza.etal_2023}
C.~Capezza, F.~Centofanti, A.~Lepore, A.~Menafoglio, B.~Palumbo, and
  S.~Vantini.
\newblock \emph{funcharts: Functional Control Charts}, 2023{\natexlab{a}}.
\newblock URL \url{https://CRAN.R-project.org/package=funcharts}.
\newblock R package version 1.3.1.

\bibitem[Capezza et~al.(2023{\natexlab{b}})Capezza, Centofanti, Lepore,
  Menafoglio, Palumbo, and Vantini]{Capezza.etal_2023c}
C.~Capezza, F.~Centofanti, A.~Lepore, A.~Menafoglio, B.~Palumbo, and
  S.~Vantini.
\newblock funcharts: control charts for multivariate functional data in {R}.
\newblock \emph{Journal of Quality Technology}, 55\penalty0 (5):\penalty0
  566–583, 2023{\natexlab{b}}.
\newblock \doi{10.1080/00224065.2023.2219012}.

\bibitem[Centofanti et~al.(2021)Centofanti, Lepore, Menafoglio, Palumbo, and
  Vantini]{Centofanti.etal_2021}
F.~Centofanti, A.~Lepore, A.~Menafoglio, B.~Palumbo, and S.~Vantini.
\newblock Functional regression control chart.
\newblock \emph{Technometrics}, 63\penalty0 (3):\penalty0 281--294, 2021.
\newblock \doi{10.1080/00401706.2020.1753581}.

\bibitem[Di et~al.(2014)Di, Crainiceanu, and Jank]{Di.etal_2014}
C.~Di, C.~M. Crainiceanu, and W.~S. Jank.
\newblock Multilevel sparse functional principal component analysis.
\newblock \emph{Stat}, 3\penalty0 (1):\penalty0 126--143, 2014.
\newblock \doi{10.1002/sta4.50}.

\bibitem[Farrar and Worden(2012)]{Farrar.Worden_2012}
C.~R. Farrar and K.~Worden.
\newblock \emph{{Structural Health Monitoring: A Machine Learning
  Perspective}}.
\newblock John Wiley \& Sons, 2012.
\newblock \doi{10.1002/9781118443118}.

\bibitem[Goldsmith et~al.(2022)Goldsmith, Scheipl, Huang, Wrobel, Di, Gellar,
  Harezlak, McLean, Swihart, Xiao, Crainiceanu, and Reiss]{Goldsmith.etal_2022}
J.~Goldsmith, F.~Scheipl, L.~Huang, J.~Wrobel, C.~Di, J.~Gellar, J.~Harezlak,
  M.~W. McLean, B.~Swihart, L.~Xiao, C.~Crainiceanu, and P.~T. Reiss.
\newblock \emph{refund: Regression with Functional Data}, 2022.
\newblock URL \url{https://CRAN.R-project.org/package=refund}.
\newblock R package version 0.1-26.

\bibitem[Han et~al.(2021)Han, Ma, Xu, and Liu]{Han.etal_2021}
Q.~Han, Q.~Ma, J.~Xu, and M.~Liu.
\newblock Structural health monitoring research under varying temperature
  condition: a review.
\newblock \emph{Journal of Civil Structural Health Monitoring}, 11:\penalty0
  149--173, 2021.
\newblock \doi{10.1007/s13349-020-00444-x}.

\bibitem[Happ and Greven(2018)]{Happ.Greven_2018}
C.~Happ and S.~Greven.
\newblock Multivariate functional principal component analysis for data
  observed on different (dimensional) domains.
\newblock \emph{Journal of the American Statistical Association}, 522\penalty0
  (113):\penalty0 649--659, 2018.
\newblock \doi{10.1080/01621459.2016.1273115}.

\bibitem[Happ-Kurz(2020)]{Happ-Kurz_2020}
C.~Happ-Kurz.
\newblock Object-oriented software for functional data.
\newblock \emph{Journal of Statistical Software}, 93\penalty0 (5):\penalty0
  1--38, 2020.
\newblock \doi{10.18637/jss.v093.i05}.

\bibitem[Happ-Kurz(2022)]{Happ-Kurz_2022}
C.~Happ-Kurz.
\newblock \emph{MFPCA: Multivariate Functional Principal Component Analysis for
  Data Observed on Different Dimensional Domains}, 2022.
\newblock URL \url{https://CRAN.R-project.org/package=MFPCA}.
\newblock R package version 1.3-10.

\bibitem[Jensen et~al.(2008)Jensen, Birch, and Woodall]{Jensen.etal_2008}
W.~A. Jensen, J.~B. Birch, and W.~H. Woodall.
\newblock Monitoring correlation within linear profiles using mixed models.
\newblock \emph{Journal of Quality Technology}, 40\penalty0 (2):\penalty0
  167--183, 2008.
\newblock \doi{10.1080/00224065.2008.11917723}.

\bibitem[Keshmiry et~al.(2023)Keshmiry, Hassani, Mousavi, and
  Dackermann]{Keshmiry.etal_2023}
A.~Keshmiry, S.~Hassani, M.~Mousavi, and U.~Dackermann.
\newblock Effects of environmental and operational conditions on structural
  health monitoring and non-destructive testing: A systematic review.
\newblock \emph{Buildings}, 13\penalty0 (4), 2023.
\newblock \doi{10.3390/buildings13040918}.

\bibitem[Koner and Staicu(2023)]{KonSta:2023}
S.~Koner and A.-M. Staicu.
\newblock Second-generation functional data.
\newblock \emph{Annual Review of Statistics and Its Application}, 10:\penalty0
  547--572, 2023.
\newblock \doi{10.1146/annurev-statistics-032921-033726}.

\bibitem[Maes and Lombaert(2020)]{Maes.Lombaert_2020}
K.~Maes and G.~Lombaert.
\newblock Monitoring railway bridge {KW51} before, during, and after
  retrofitting. v1.0.
\newblock 2020.
\newblock \doi{10.5281/zenodo.3745914}.

\bibitem[Maes and Lombaert(2021)]{Maes.Lombaert_2021}
K.~Maes and G.~Lombaert.
\newblock Monitoring railway bridge {KW51} before, during, and after
  retrofitting.
\newblock \emph{Journal of Bridge Engineering}, 26\penalty0 (3):\penalty0
  04721001, 2021.
\newblock \doi{10.1061/(ASCE)BE.1943-5592.0001668}.

\bibitem[Maes et~al.(2022)Maes, {Van Meerbeeck}, Reynders, and
  Lombaert]{Maes.etal_2022}
K.~Maes, L.~{Van Meerbeeck}, E.~P. Reynders, and G.~Lombaert.
\newblock Validation of vibration-based structural health monitoring on
  retrofitted railway bridge {KW51}.
\newblock \emph{Mechanical Systems and Signal Processing}, 165:\penalty0
  108380, 2022.
\newblock \doi{10.1016/j.ymssp.2021.108380}.

\bibitem[Momeni and Ebrahimkhanlou(2022)]{Momeni.Ebrahimkhanlou_2022}
H.~Momeni and A.~Ebrahimkhanlou.
\newblock High-dimensional data analytics in structural health monitoring and
  non-destructive evaluation: a review paper.
\newblock \emph{Smart Materials and Structures}, 31\penalty0 (4):\penalty0
  043001, 2022.
\newblock \doi{10.1088/1361-665X/ac50f4}.

\bibitem[Pourhoseingholi et~al.(2012)Pourhoseingholi, Baghestani, and
  Vahed]{Pourhoseingholi.etal_2012}
M.~A. Pourhoseingholi, A.~R. Baghestani, and M.~Vahed.
\newblock How to control confounding effects by statistical analysis.
\newblock \emph{Gastroenterology and Hepatology From Bed to Bench}, 5\penalty0
  (2):\penalty0 79--83, 2012.
\newblock \doi{10.22037/ghfbb.v5i2.246}.

\bibitem[{R Core Team}(2023)]{R_2023}
{R Core Team}.
\newblock \emph{{R: A Language and Environment for Statistical Computing}}.
\newblock R Foundation for Statistical Computing, Vienna, Austria, 2023.
\newblock URL \url{https://www.R-project.org/}.

\bibitem[Rainieri and Fabbrocino(2014)]{Rainieri.Fabbrocino_2014}
C.~Rainieri and G.~Fabbrocino.
\newblock \emph{{Operational Modal Analysis of Civil Engineering Structures}}.
\newblock Springer New York, NY, 2014.
\newblock \doi{10.1007/978-1-4939-0767-0}.

\bibitem[Ramsay and Silverman(2005)]{Ramsay.Silvermann_2005}
J.~O. Ramsay and B.~W. Silverman.
\newblock \emph{Functional Data Analysis}.
\newblock Springer New York, 2005.
\newblock \doi{10.1007/b98888}.

\bibitem[Ramsay et~al.(2022)Ramsay, Graves, and Hooker]{Ramsey.etal_2022}
J.~O. Ramsay, S.~Graves, and G.~Hooker.
\newblock \emph{fda: Functional Data Analysis}, 2022.
\newblock URL \url{https://CRAN.R-project.org/package=fda}.
\newblock R package version 6.0.5.

\bibitem[Scheipl et~al.(2016)Scheipl, Gertheiss, and Greven]{Scheipl.etal_2016}
F.~Scheipl, J.~Gertheiss, and S.~Greven.
\newblock {Generalized functional additive mixed models}.
\newblock \emph{Electronic Journal of Statistics}, 10\penalty0 (1):\penalty0
  1455 -- 1492, 2016.
\newblock \doi{10.1214/16-EJS1145}.

\bibitem[Sievert(2020)]{Sievert_2020}
C.~Sievert.
\newblock \emph{Interactive Web-Based Data Visualization with R, plotly, and
  shiny}.
\newblock Chapman and Hall/CRC, 2020.
\newblock \doi{10.1201/9780429447273}.

\bibitem[Sohn et~al.(2000)Sohn, Czarnecki, and Farrar]{Sohn:Czar:Farr:2000}
H.~Sohn, J.~A. Czarnecki, and C.~R. Farrar.
\newblock Structural health monitoring using statistical process control.
\newblock \emph{Journal of Structural Engineering}, 126:\penalty0 1356--1363,
  2000.
\newblock \doi{10.1061/(ASCE)0733-9445(2000)126:11(1356)}.

\bibitem[Xiao et~al.(2013)Xiao, Li, and Ruppert]{Xiao.etal_2013}
L.~Xiao, Y.~Li, and D.~Ruppert.
\newblock Fast bivariate p-splines: the sandwich smoother.
\newblock \emph{Journal of the Royal Statistical Society: Series B},
  75\penalty0 (3):\penalty0 577--599, 2013.
\newblock \doi{10.1111/rssb.12007}.

\bibitem[Yao et~al.(2005)Yao, Müller, and Wang]{Yao.etal_2005}
F.~Yao, H.-G. Müller, and J.-L. Wang.
\newblock Functional data analysis for sparse longitudinal data.
\newblock \emph{Journal of the American Statistical Association}, 100\penalty0
  (470):\penalty0 577--590, 2005.
\newblock \doi{10.1198/016214504000001745}.

\end{thebibliography}

\end{document}